\documentclass[aps,prb,groupedaddress,showpacs,floatfix]{revtex4}
\usepackage{graphicx}
\begin{document}

\title{Influence of contacts on the microwave response of a two-dimensional electron stripe}

\author{S. A. Mikhailov}
\email[E-mail address: ]{sergey.mikhailov@miun.se}
\affiliation{Mid-Sweden University, ITM, Electronics Design Division, 851 70 Sundsvall, Sweden}
\author{N. A. Savostianova}
\affiliation{Mid-Sweden University, ITM, Electronics Design Division, 851 70 Sundsvall, Sweden}

\date{\today}

\begin{abstract}
Electromagnetic response of a finite-width two-dimensional electron stripe with attached metallic side contacts is theoretically studied. It is shown that contacts substantially influence the position, the linewidth, and the amplitude of plasmon-polariton resonances in the stripe. In finite magnetic fields, absorption of the wave with the inactive circular polarization (which is not absorbed in an infinite system without contacts) may become larger than that of the wave with the active polarization. The results are discussed in view of recent microwave experiments in two-dimensional electron systems.
\end{abstract}

\pacs{73.20.Mf, 73.21.-b, 71.36.+c, 78.70.Gq}

\maketitle

\section{\label{intro}Introduction}

Recently, a plenty of experimental and theoretical work has been done on studying two-dimensional (2D) electron systems under microwave irradiation. The giant $1/B$-periodic magnetoresistance oscillations \cite{Zudov01,Ye01} and zero resistance states \cite{Mani02,Zudov03,Dorozhkin03,Yang03,Mani04,Studenikin04,Willett04,Du04,Smet05} (ZRS) have been discovered at $\omega\gtrsim\omega_c$, and new $B$-periodic photoresistance and photovoltage oscillations, surviving at up to liquid nitrogen temperatures\cite{Kukushkin04a,Mikhailov-SPIE04,Kukushkin05a}, have been found at $\omega\lesssim\omega_c$ (here $\omega$ and $\omega_c$ are the microwave and the cyclotron frequencies). Both phenomena are not yet fully understood. Theoretical scenarios, developed for explanation of ZRS (e.g. \cite{Durst03,Andreev03,Shi03,Koulakov03,Dmitriev03,Dmitriev04,Dmitriev05,Lei03a,Ryzhii70,Ryzhii86,Ryzhii03a,Ryzhii03b,Ryzhii04,Ryzhii04a,Ryzhii04b,Ryzhii05}), did not explain the plasma-shift paradox \cite{Mikhailov03c,Mikhailov04a} and turned out to be in evident contradiction with results of the recent experiment \cite{Smet05}, where independence of the effect on the sense of circular polarization has been proved (see discussion in Ref. \cite{Smet05}). $B$-periodic oscillations have been explained \cite{Kukushkin04a} in terms of excitation and interference of edge magnetoplasmons \cite{Volkov88,Mikhailov00a} in the system, however, a more detailed treatment of this phenomenon is also needed, especially in view of its possible applications for giga- and terahertz detection and spectroscopy of radiation \cite{Kukushkin05a}.

It has been demonstrated both experimentally\cite{Kukushkin03a,Kukushkin03b} and theoretically\cite{Mikhailov04a,Mikhailov05a}, that a new physics becomes essential in macroscopic samples at low (microwave) frequencies. Radiative and retardation effects, which could be ignored at higher (far-infrared) frequencies, are important and should be included in the theory at microwaves. The goal of this paper is to investigate the {\em influence of contacts} on the microwave response of a 2D electron-gas (EG) stripe. We will show that, in the presence of well conducting contacts, microwave response of the system should be calculated taking into account retardation and radiative effects even if the response of the same system without contacts were sufficient to study in the quasi-static approximation. In general, the role of contacts can be important in both types of experiments mentioned above. Indeed, the size of the contact pads in the experiments on Hall bar samples (0.1 -- 1 mm) is comparable to the lateral dimensions of the samples and the wavelength of radiation. In the Corbino-disk experiments\cite{Yang03} the size of contacts is even larger. Independence of the ZRS effect on the sense of circular polarization \cite{Smet05} may indicate that this is not a bulk but an edge effect, related to the boundary of the 2D gas with the contact regions (that the zero resistance states can be due not to bulk but to edge effects was first pointed out in Ref. \cite{Mikhailov03c}). In the effect of $B$-periodic oscillations the role of contacts also needs to be better understood, as it is due to contacts the edge magnetoplasmons are excited in the sample \cite{Kukushkin04a,Mikhailov-SPIE04,Kukushkin05a}.

Plasma oscillations in a 2D EG stripe with metallic side contacts have been recently studied in Ref. \cite{Satou05}. In \cite{Satou05}, the positions and the damping rates of 2D plasmons in zero external magnetic field have been found in quasi-static approximation. We study the problem taking into account electrodynamic effects \cite{Mikhailov04a} and investigate the system response in the presence of a magnetic field. 

\section{\label{form}Formalism}

\begin{figure}
\includegraphics[width=8.5cm]{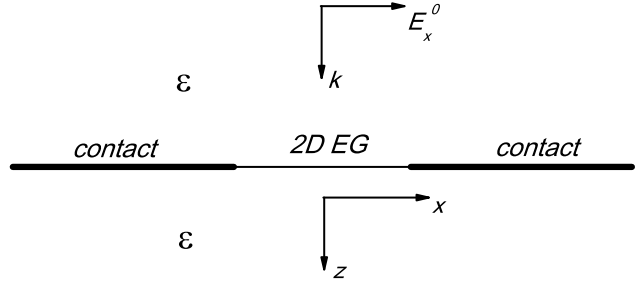}
\caption{Geometry of the considered structure. The $y$ axis is directed toward the reader, and the system is infinite in the $y$ direction. The incident electromagnetic wave with the wavevector $k=k_z$ may also have the $E_y^0$ component of the electric field. }
\label{geometry}
\end{figure}

Consider a 2D EG stripe lying in the plane $z=0$ and occupying the area $|x|<W/2$, $-\infty<y<\infty$, Figure \ref{geometry}. The contacts are modelled by thin (two-dimensional) layers with the surface conductivity $\sigma_c$ lying in the same plane $z=0$ at $|x|>W/2$. Possible nonlinear (Schottky) effects at the boundary 2D-gas -- contacts are neglected. The system is placed in a magnetic field ${\bf B}=(0,0,B)$, and the dielectric constant $\epsilon$ of surrounding medium is assumed to be uniform in all the space. 

If an external plane electromagnetic wave, with the frequency $\omega$ and the electric field ${\bf E}^{ext}({\bf r},t)={\bf E}^{0}e^{i\omega(z\sqrt{\epsilon}/c- t)}$, is normally incident upon the structure, the total electric field ${\bf E}^{tot}={\bf E}^{ext}+{\bf E}^{ind}$ is determined by the Maxwell equations 
\begin{equation}
\nabla \times (\nabla \times {\bf E}^{tot}) - (\omega^2\epsilon/c^2)
{\bf E}^{tot} = \nabla \times (\nabla \times {\bf E}^{ind}) - (\omega^2\epsilon/c^2)
{\bf E}^{ind} =  (4\pi i\omega/c^2)
{\bf j}(x,\omega)\delta(z),
\label{me}
\end{equation} 
where ${\bf E}^{ind}$ is the induced field of the scattered wave. The current ${\bf j}(x,\omega)$ in (\ref{me}) is given by the relation $j_\alpha(x,\omega)=\sigma_{\alpha\beta}(x,\omega)E_{\beta}^{tot}(x,\omega)$, where the local 2D conductivity of the system is
\begin{equation}
\sigma_{\alpha\beta}(x,\omega)=\sigma_{\alpha\beta}(\omega)\theta(W/2-|x|)+\sigma_c\delta_{\alpha\beta}\theta(|x|-W/2).
\label{sigma}
\end{equation}
Here $\theta(x)$ is the step function, and $\sigma_{\alpha\beta}(\omega)$ is the conductivity of the 2D EG. In (\ref{sigma}) we assume that $\sigma_{\alpha\beta}(\omega)$ is a function of the frequency $\omega$, the magnetic field $B$, the 2D EG density $n_s$, and the momentum relaxation rate $\gamma$, while the conductivity of contacts $\sigma_c$ is considered to be a constant. This corresponds to the experimentally relevant situation when the momentum relaxation rate of electrons in contacts exceeds both their cyclotron frequency and the frequency of radiation.

Assuming that all physical quantities do not depend on the $y$-coordinate, and looking for a solution in the form
\begin{equation}
{\bf E}^{ind}(x,z,\omega)=\int dq_x{\bf E}^{ind}(q_x,z,\omega)e^{iq_xx},
\label{look}
\end{equation}
we get the following relations between Fourier components of the electric fields and currents,
\begin{equation}
\frac{\partial^2E_x^{ind}(q_x,z)}{\partial z^2}-\kappa^2E_x^{ind}(q_x,z)=\frac{4\pi i\sigma_c\kappa^2}{\omega\epsilon}E_x^{tot}(q_x,0)\delta(z)+\frac{4\pi i\kappa^2}{\omega\epsilon}\delta j_x(q_x)\delta(z),
\label{eqEx}
\end{equation}
\begin{equation}
\frac{\partial^2E_y^{ind}(q_x,z)}{\partial z^2}-\kappa^2E_y^{ind}(q_x,z)=-\frac{4\pi i\omega\sigma_c}{c^2}E_y^{tot}(q_x,0)\delta(z)-\frac{4\pi i\omega}{c^2}\delta j_y(q_x)\delta(z).
\label{eqEy}
\end{equation}
Here $\delta j_\alpha=(\sigma_{\alpha\beta}-\sigma_c\delta_{\alpha\beta})\theta(W/2-|x|)E_\beta^{tot}$ and $\kappa^2=q_x^2-\omega^2\epsilon/c^2$. Equations (\ref{eqEx}) and (\ref{eqEy}) allows us to relate the induced field at all $z$ to the same field at the plane $z=0$, 
\begin{equation}
E_{x,y}^{ind}(q_x,z)=E_{x,y}^{ind}(q_x,0)e^{-\kappa|z|},
\label{i}
\end{equation}
and the field at $z=0$ to the current $\delta j_\alpha$,
\begin{equation}
E_x^{ind}(q_x,0)=
-\frac
{2\pi i\kappa}
{\omega\epsilon+2\pi i\kappa\sigma_c}\left[ \sigma_c E_x^{ext}(q_x,0) + \delta j_x(q_x)\right],
\label{Exinq}
\end{equation}
\begin{equation}
E_y^{ind}(q_x,0)=
\frac
{2\pi i\omega}
{c^2\kappa-2\pi i\omega\sigma_c}\left[ \sigma_c E_y^{ext}(q_x,0) + \delta j_y(q_x)\right];
\label{Eyinq}
\end{equation}
here $E_{x,y}^{ext}(q_x,0)=E_{x,y}^{0}\delta(q_x)$. 
The $z$-dependence of the field (\ref{look}) is determined by the multiplier $e^{-\kappa|z|}$ in (\ref{i}). Parts of the integral (\ref{look}), corresponding to $q_x^2>\omega^2\epsilon/c^2$ and $q_x^2<\omega^2\epsilon/c^2$, describe the evanescent induced fields ($\kappa$ is real) and the scattered outgoing wave ($\kappa$ is imaginary), respectively. The sign of Im $\kappa$ in the latter case should be chosen in accordance with the scattering boundary conditions at $|z|\to\infty$ ($\kappa=-i\sqrt{\omega^2\epsilon/c^2-q_x^2}$ at $q_x^2<\omega^2\epsilon/c^2$).

If $\sigma_c=0$, the problem (\ref{Exinq}) -- (\ref{Eyinq}) is reduced to the one solved in Ref. \cite{Mikhailov05a}. If $\sigma_c\neq 0$, Eqs. (\ref{Exinq}) and (\ref{Eyinq}) lead to the following system of integral equations for the fields at the plane $z=0$
\begin{eqnarray}
E_x^{tot}(x)&=&\frac{E_x^{0}}{1+\eta_c}
+\left(1-\frac{\sigma_{xx}}{\sigma_c}\right)E_x^{tot}(x)\Theta(W/2-|x|)-\frac {\omega\epsilon}{2\pi}\left(1-\frac{\sigma_{xx}}{\sigma_c}\right)\int\frac{dqe^{iqx}}{\omega\epsilon+2\pi i\kappa\sigma_c}\int_{-W/2}^{W/2}dx'e^{-iqx'}E_x^{tot}(x')
\nonumber \\
&-&
\frac{\sigma_{xy}}{\sigma_c}E_y^{tot}(x)\Theta(W/2-|x|)+\frac {\omega\epsilon}{2\pi}\frac{\sigma_{xy}}{\sigma_c}\int\frac{dqe^{iqx}}{\omega\epsilon+2\pi i\kappa\sigma_c}\int_{-W/2}^{W/2}dx'e^{-iqx'}E_y^{tot}(x'),
\label{ex}
\end{eqnarray}
\begin{eqnarray}
E_y^{tot}(x)&=&\frac{E_y^{0}}{1+\eta_c}
+i\omega\sigma_{yx}\int\frac{dqe^{iqx}}{c^2\kappa-2\pi i\omega\sigma_c}\int_{-W/2}^{W/2}dx'e^{-iqx'}E_x^{tot}(x')
\nonumber \\
&+&
i\omega(\sigma_{yy}-\sigma_c)\int \frac{dqe^{iqx}}{c^2\kappa-2\pi i\omega\sigma_c}\int_{-W/2}^{W/2}dx'e^{-iqx'}E_y^{tot}(x').
\label{ey}
\end{eqnarray}
The dimensionless parameter 
\begin{equation}
\eta_c=2\pi \sigma_c/c\sqrt{\epsilon}
\end{equation}
here characterizes the conductivity of the contact areas; for a good metal, typically, $\eta_c\gg 1$.

Equations (\ref{ex}) and (\ref{ey}) are valid at $-\infty<x<+\infty$. Inside the stripe ($|x|<W/2$) Eq. (\ref{ex}) can be rewritten as
\begin{eqnarray}
E_x^{tot}(x)&=&\frac{\sigma_c}{\sigma_{xx}}\frac{E_x^{0}}{1+\eta_c}-\frac {\omega\epsilon}{2\pi}\left(\frac{\sigma_c}{\sigma_{xx}}-1\right)\int\frac{dqe^{iqx}}{\omega\epsilon+2\pi i\kappa\sigma_c}\int_{-W/2}^{W/2}dx'e^{-iqx'}E_x^{tot}(x')
\nonumber \\
&-&
\frac{\sigma_{xy}}{\sigma_{xx}}E_y^{tot}(x)+\frac {\omega\epsilon}{2\pi}\frac{\sigma_{xy}}{\sigma_{xx}}\int\frac{dqe^{iqx}}{\omega\epsilon+2\pi i\kappa\sigma_c}\int_{-W/2}^{W/2}dx'e^{-iqx'}E_y^{tot}(x').
\label{ex'}
\end{eqnarray}
To solve the system of equations (\ref{ex'}) and (\ref{ey}) at $|x|<W/2$, we expand the fields 
\begin{equation}
\left(
\begin{array}{c}
E_x^{tot}(x)\\
E_y^{tot}(x)
\end{array}\right)
=\sum_{n=0}
\left(
\begin{array}{c}
A_n\\
B_n
\end{array}\right)
\cos\frac{2\pi xn}W,
\label{Exy}
\end{equation}
and reduce the problem to the set of matrix equations
\begin{equation}
\sum_{n=0}\left\{{\cal E}_{mn}+\left(Z-1\right){\cal K}_{mn}\right\}A_n+Z_c
\sum_{n=0}\left\{{\cal E}_{mn}-{\cal K}_{mn}\right\}B_n=\delta_{m0}ZE_x^0/(1+\eta_c),
\label{me1}
\end{equation}
\begin{equation}
-Z_c
\sum_{n=0}{\cal L}_{mn}A_n+
\sum_{n=0}\left\{Z{\cal E}_{mn}-\left(Z-1\right){\cal L}_{mn}\right\}B_n=\delta_{m0}ZE_y^0/(1+\eta_c).
\label{me2}
\end{equation}
Here 
\begin{equation}
{\cal E}_{mn}=\frac 12\delta_{mn}(1+\delta_{m0}),
\end{equation}
\begin{equation}
{\cal K}_{mn}=2\int_0^\infty \frac{dQ F_m(Q)F_n(Q)}{1+\frac{i\eta_c}\Omega \sqrt{Q^2-\Omega^2}},
\end{equation}
\begin{equation}
{\cal L}_{mn}=2\int_0^\infty \frac{dQ F_m(Q)F_n(Q)}{1+\frac{i}{\eta_c\Omega}\sqrt{Q^2-\Omega^2}},
\end{equation}
$\Omega= \omega W\sqrt{\epsilon}/2\pi c$, $Z=\sigma_c/\sigma_{xx}=\sigma_c/\sigma_{yy}$, $Z_c=\sigma_{xy}/\sigma_{xx}=-\sigma_{yx}/\sigma_{xx}$, $Q=qW/2\pi$, and
\begin{equation}
F_n(Q)=
\frac 12\left\{\frac{\sin\pi(Q-n)}{\pi(Q-n)}+\frac{\sin\pi(Q+n)}{\pi(Q+n)}\right\}.
\end{equation}
Notice that the matrix elements ${\cal K}_{mn}$ and ${\cal L}_{mn}$ depend only on the properties of contacts and are not influenced by parameters of the 2D EG (density, mobility, etc). They are complex, which physically corresponds to emission of waves from the stripe and leads to the radiative decay of plasma modes. The dimensionless frequency $\Omega$ is the ratio of the stripe width $W$ to the wavelength of radiation $\lambda$. 

Having solved equations (\ref{me1})--(\ref{me2}), we get the coefficients $A_n$, $B_n$ and the electric field (\ref{Exy}). Then, we calculate the Joule heat inside the stripe $J=\int_{-W/2}^{W/2}dx [{\bf j}^*{\bf E}^{tot}+{\bf j}({\bf E}^{tot})^*]/4$ and the absorption coefficient defined as the ratio of $J$ to the energy flow incident on the stripe
\begin{equation}
A=\frac{8\pi J}{cW\sqrt{\epsilon}(|E_x^0|^2+|E_y^0|^2)}.
\label{absor}
\end{equation}
In the next Section we show $A$ as a function of frequency, magnetic field and other parameters of the system. 

\section{\label{res}results and discussions}

We measure the frequency, the magnetic field, and the momentum relaxation rate $\gamma=e/m^\star\mu$ in units $\omega/\omega_0$, $\omega_c/\omega_0$, and $\gamma/\omega_0$, where
\begin{equation}
\omega_0=\sqrt{\frac{2\pi^2 n_se^2}{m^\star\epsilon W}}
\label{omega0}
\end{equation}
is the frequency of the quasi-static 2D plasmon with the wavector $q= \pi/W$, $\mu$, $e$, and $m^\star$ are the mobility, the charge, and the effective mass of 2D electrons, respectively. The retardation parameter of the 2D EG
\begin{equation}
\alpha=\frac{\omega_0 \sqrt{\epsilon}W}{\pi c}=\sqrt{\frac{2n_se^2W}{m^\star c^2}}=\frac\Gamma{\omega_0}
\label{alpha}
\end{equation}
is defined\cite{Mikhailov05a} as the ratio of the frequency $\omega_0$ to the frequency of light $cq/\sqrt{\epsilon}$ with the same wavevector $q= \pi/W$. In a stripe without contacts small $\alpha$'s correspond to the quasistatic limit, while at $\alpha\simeq 1$ electrodynamic effects become important \cite{Mikhailov05a}. In this paper we mainly analyze the case $\alpha\ll 1$: it will be shown that in systems with well conducting contacts electrodynamic effects are important even if $\alpha$ is small. In (\ref{alpha}), the frequency 
\begin{equation}
\Gamma=\frac{2\pi n_se^2}{m^\star c\sqrt{\epsilon}}
\end{equation}
is the radiative decay rate of an infinite uniform 2D electron layer\cite{Chiu76,Matov92,Matov96,Mikhailov96a,Popov98,Popov01,Mikhailov04a}. For $\sigma_{\alpha\beta}(\omega)$ we will use the Drude formulas. 

\subsection{\label{zeroB}Zero magnetic field}

\begin{figure}
\includegraphics[width=8.5cm]{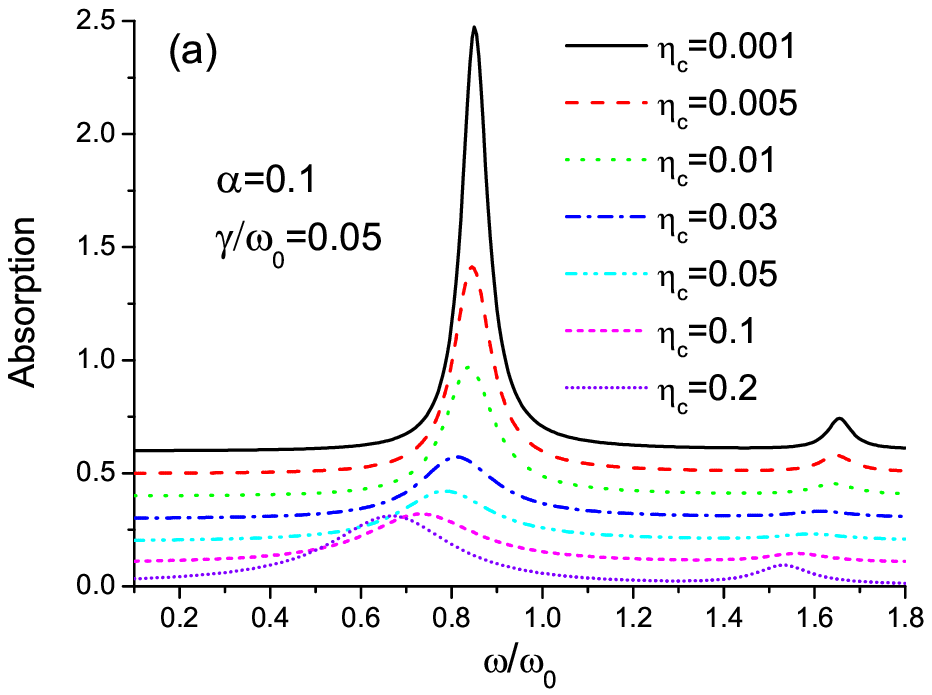}
\includegraphics[width=8.5cm]{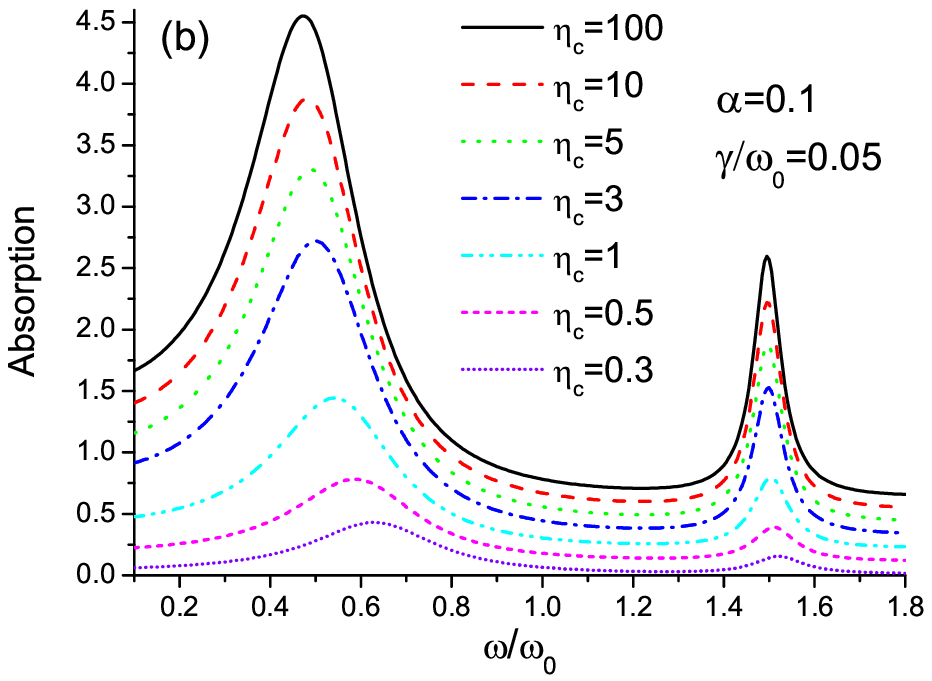}\\
\includegraphics[width=8.5cm]{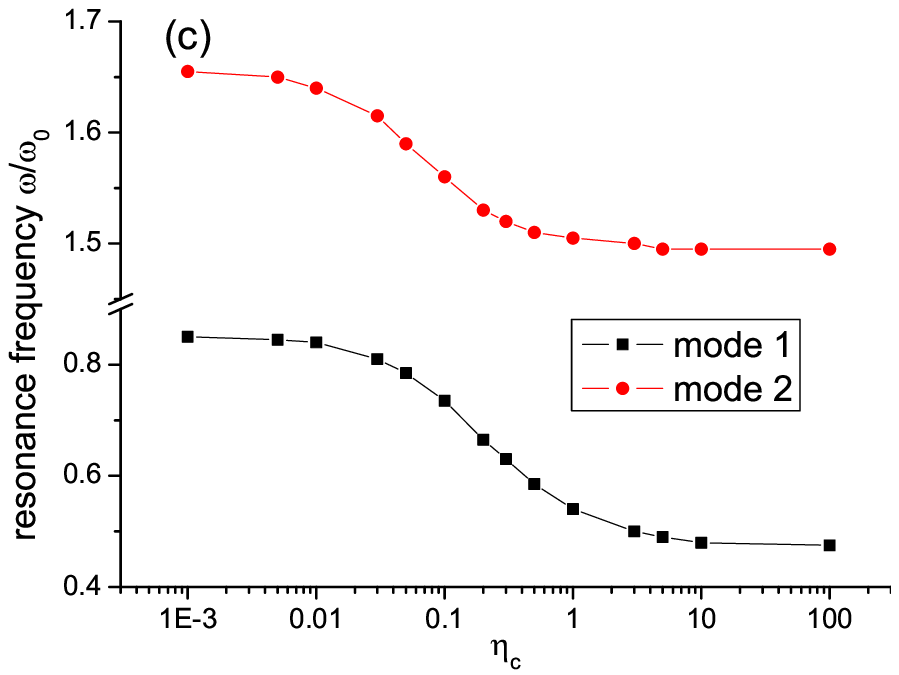}
\includegraphics[width=8.5cm]{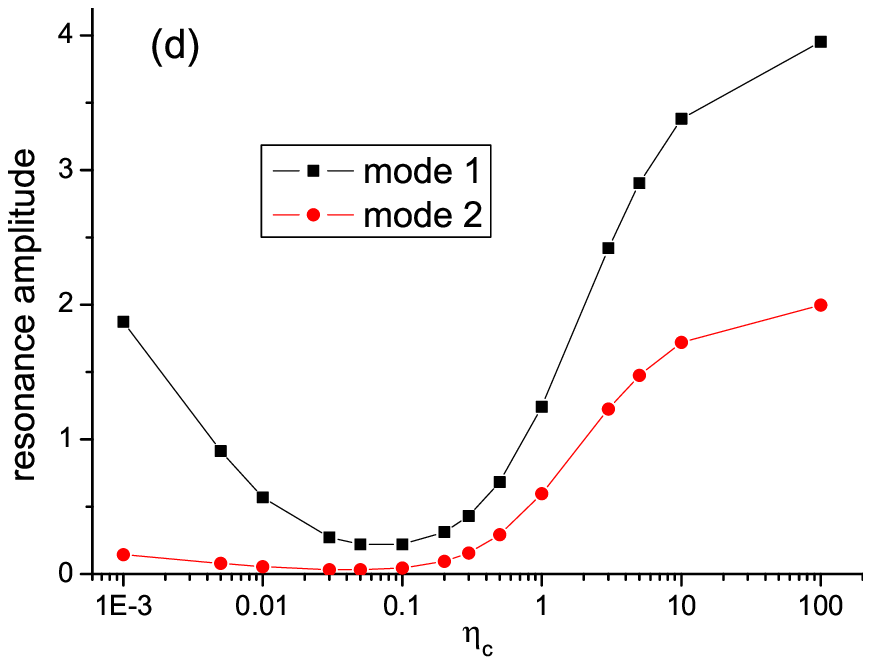}
\caption{(Color online) (a) and (b): Absorption spectra of a 2D EG stripe placed between contacts at $\gamma/\omega_0=0.05$, $\alpha=0.1$, and different values of the parameter $\eta_c$. The absorption curves are vertically shifted by 0.1 for clarity. The first and second resonance modes are shown (compare to Figure 1 in Ref. \cite{Mikhailov05a}). The position (c) and the amplitude (d) of resonances as a function of $\eta_c$. }
\label{B0a0.1}
\end{figure}

First, we analyze absorption spectra of a 2D EG stripe in the absence of magnetic field and at a small value of the retardation parameter $\alpha=0.1$, Figure \ref{B0a0.1}. The incident electromagnetic wave is assumed to be linearly polarized along the $x$-axis (perpendicular to the stripe). At small values of $\eta_c$ ($\lesssim 0.001$) the absorption spectrum is very close to that of the stripe without contacts (compare to Figure 1 from Ref. \cite{Mikhailov05a}). When $\eta_c$ increases, the both resonances first experience a red shift and essential reduction of the amplitudes, Figure \ref{B0a0.1}a. Then the both resonances grow up in amplitude again, Figure \ref{B0a0.1}b, with their positions remaining essentially the same. Both modes show a distinct broadening and reduction of the quality factor with the growing $\eta_c$. The relative height of the first and the second modes also substantially changes with $\eta_c$: while at small $\eta_c$ the second mode is much smaller in amplitude than the first one, at $\eta_c\gtrsim 1$ they are comparable in amplitude, Figure \ref{B0a0.1}d. The shift of resonances to lower frequencies, quantitatively illustrated in Figure \ref{B0a0.1}c, is in agreement with similar results of Ref. \cite{Satou05} (Figure 5 there). The linewidth of resonances is however essentially larger in our calculations as compared to Ref. \cite{Satou05} (Figure 6 there). The reason is that the radiative decay which has been included in our scheme, has been ignored in the formalism of Ref. \cite{Satou05}.

Physically, the broadening of modes and the variation of their amplitudes can be explained as follows. In the absence of contacts\cite{Mikhailov05a}, the incident radiation induces oscillating charge lines near the stripe boundaries and excites 2D plasmons in it. The linewidth of these plasmon resonances is determined by the energy losses due to scattering of electrons in the stripe and by reradiation of waves by oscillating charges  \cite{Mikhailov05a}. When the contact wings are attached to the stripe and the conductivity of metal becomes comparable with that of the 2DEG ($\sigma_c\simeq|\sigma_{xx}(\omega)|$, or $\eta_c\simeq\alpha$), the plasmon charge leaks out to contacts and the amplitude of resonances becomes minimal, Figure \ref{B0a0.1}d. At even higher contact conductivities, the value of oscillating charges, induced near the edges of the stripe, becomes very large, as they are determined by big currents flowing to the boundary from the highly conducting metal sides. In other words, the well conducting contact wings serve as antennas,\cite{Balanis05} collecting the incident radiation energy from a larger area and supplying it to the 2D EG. This leads to the growth of resonance amplitudes. On the other hand, the linewidth of resonances becomes bigger, and the quality factor -- lower, because attaching good contacts to the stripe increases the dipole radiation of plasmons and hence their radiative decay. The red shift of modes is caused by screening of the plasmon field by electrons in contacts. 

It should be noticed that, quantitatively, the value of the radiative decay and the amplitude of resonances depend on the dielectric properties of the surrounding medium and, in particular, on the shape and dimensions of contacts. In our model (Figure \ref{geometry}) the contacts are assumed to be infinite. This means that our results will reasonable well describe the real situation, if the size of the contact wings $W_c$ exceeds the wavelength of radiation. If this is not so, one should expect a reduced effect of contacts at $\lambda\simeq W_c$ and no influence at $W_c/\lambda\to 0$. At $W_c/\lambda\lesssim 1$ response of the system should also depend on the size and shape of contact wings, -- the issue considered in the theory of antennas \cite{Balanis05}, but this is already beyond the scope of our work. 

\begin{figure}
\includegraphics[width=8.5cm]{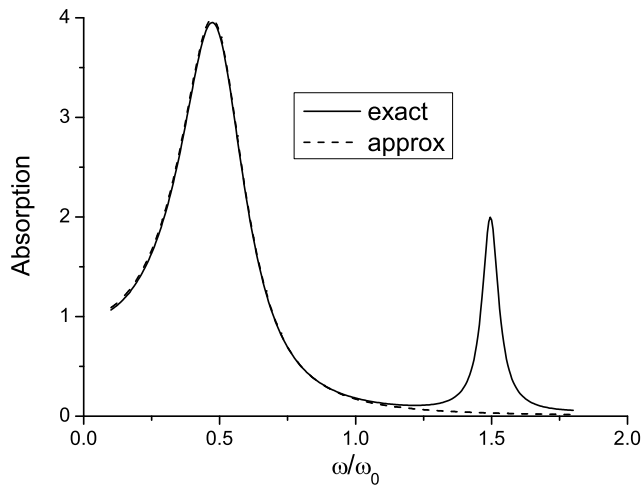}
\caption{Comparison of exact and approximate absorption spectra of a 2D EG stripe, placed between contacts: solid curve -- the exact solution, calculated with $N_h=8$ (convergency has been achieved), dashed curve -- the analytical expression (\ref{Abs0}). Parameters used are: $\gamma/\omega_0=0.05$, $\alpha=0.1$, $\eta_c=100$. }
\label{comparison}
\end{figure}

Consider now the practically important case of the very large conductivity of contacts ($\eta_c\gg 1$, $\sigma_c\gg|\sigma_{xx}|$) in more detail. In this limit, the shape of the first absorption maximum can be described analytically. Taking into account only the first term in the expansions (\ref{Exy}), 
\begin{equation}
E_x^{tot}(x)\approx A_0 =\overline{E_x^{tot}}
\end{equation}
(here $\overline{E_x^{tot}}$ is the field averaged over the width of the stripe), we get from Eqs. (\ref{me1}) and (\ref{absor}) 
\begin{equation}
\frac{\overline{E_x^{tot}}}{E_x^0}  =
\frac{Z}{(1+\eta_c)[1+(Z-1){\cal K}_{00}]},
\label{field0}
\end{equation}
\begin{equation}
A\approx 
\frac{4\pi\sigma'_{xx}}{c\sqrt{\epsilon}}
\left| 
\frac{\overline{E_x^{tot}}}{E_x^0} 
\right|^2.
\label{A0}
\end{equation}
If $Z=1$, when the stripe consists of the same material as the contact regions, $\sigma_{xx}=\sigma_c$, Eqs. (\ref{field0}) and (\ref{A0}) reproduce the known results $\overline{E_x^{tot}}/E_x^0=1/(1+\eta_c)$ and $A=2\eta_c/(1+\eta_c)$, which can be directly obtained \cite{Chiu76} by considering scattering of electromagnetic wave on a conducting sheet with the conductivity $\sigma_c$. 
If $\Omega\ll 1$ the function ${\cal K}_{00}$ is estimated as (see Appendix \ref{ap1})
\begin{equation}
{\cal K}_{00}\approx
\frac{2\Omega}{i\eta_c}
\ln\frac {0.8 i}{\Omega}
,
\label{K00a}
\end{equation}
and we get
\begin{equation}
A\approx 
\frac{2\gamma\Gamma}{\left|\Gamma-\omega(\omega+i\gamma)\frac{\alpha}{\omega_0}\left(\ln\frac{0.8}{\Omega}+i\frac\pi 2\right)\right|^2}
\label{Abs0}
\end{equation}
(the condition $\Omega\ll 1$ is satisfied, because $\Omega\equiv\omega W\sqrt{\epsilon}/2\pi c=\omega\alpha/2\omega_0$ and we consider the case $\alpha\ll 1$ and $\omega/\omega_0\simeq 1$). Figure \ref{comparison} illustrates the accuracy of Eq. (\ref{Abs0}). One sees that exact and approximate curves practically coinside around the first resonance. The second resonance cannot be, evidently, described by the simple formula (\ref{Abs0}), as it has been obtained by neglecting all but one terms in the expansion (\ref{Exy}).

\begin{figure}
\includegraphics[width=8.5cm]{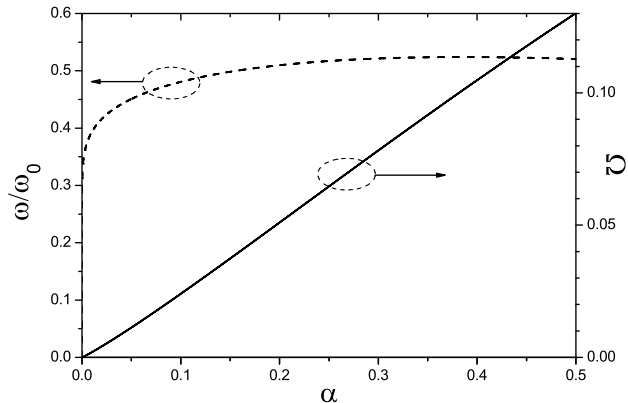}
\caption{The position of the first resonance as a function of $\alpha$, calculated from the maximum of the curve (\ref{Abs1}). The frequency units $\omega/\omega_0$ and $\Omega$ are used on the left and right axes, respectively.}
\label{resonance}
\end{figure}

As seen from Eq. (\ref{Abs0}), two factors determine the linewidth of the absorption resonance. The first factor ($i\gamma$) is due to the scattering of electrons in the 2D EG, the second one ($i\pi/2$) comes from the matrix element ${\cal K}_{00}$ and describes the radiative decay of plasma oscillations due to the presence of the contact wings (see Appendix \ref{ap1}). In 2D electron systems of high quality the first factor can be neglected, and Eq. (\ref{Abs0}) can finally be written as
\begin{equation}
A\approx 
\frac{2\gamma}{\omega_0}
\frac{\alpha^3}{\left(\alpha^2-4\Omega^2\ln\frac{0.8}{\Omega}\right)^2+\left(2\pi\Omega^2\right)^2}.
\label{Abs1}
\end{equation}
Figure \ref{resonance} shows the dependence of the resonance position, calculated from (\ref{Abs1}), as a function of the retardation parameter $\alpha$. 

\begin{figure}
\includegraphics[width=8.5cm]{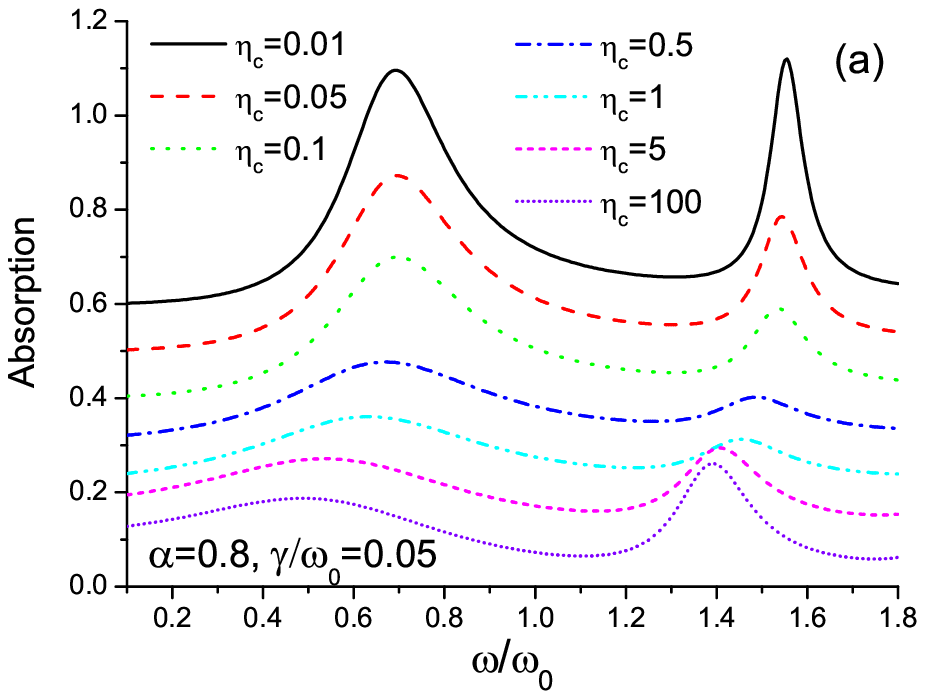}
\includegraphics[width=8.5cm]{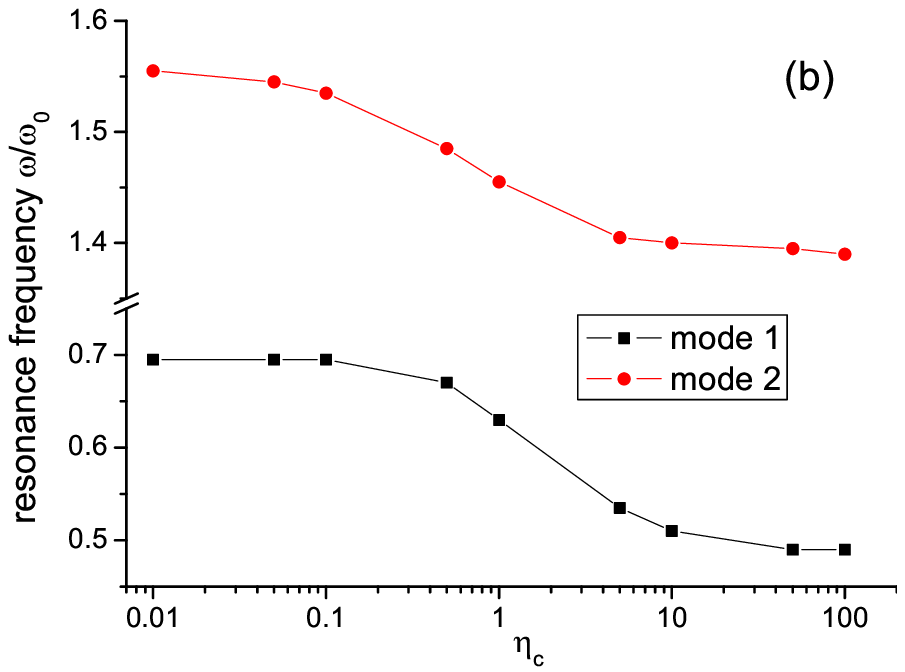}\\
\includegraphics[width=8.5cm]{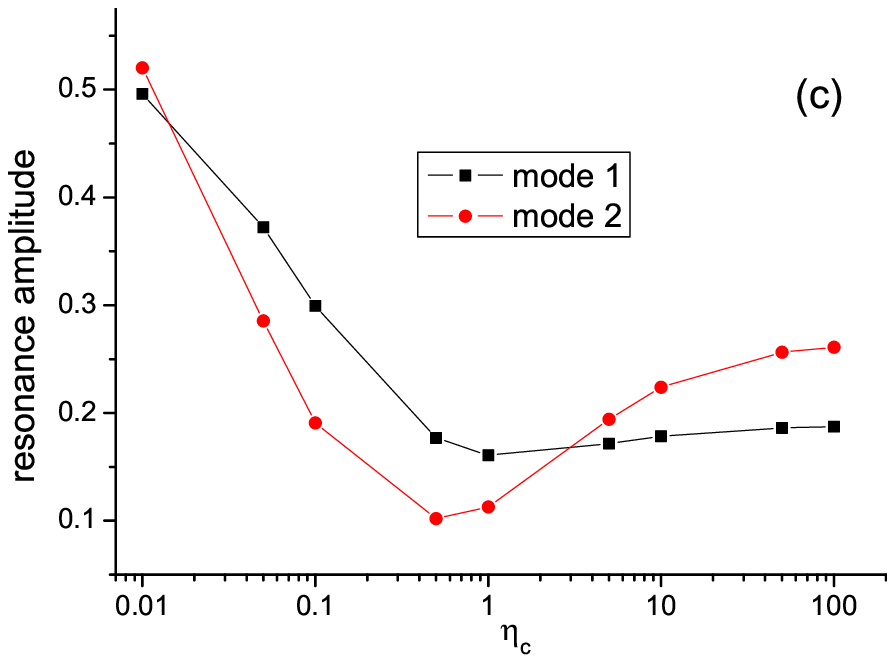}
\caption{(Color online) (a) Absorption spectra of a 2D EG stripe placed between contacts at $\gamma/\omega_0=0.05$, $\alpha=0.8$, and different values of the parameter $\eta_c$. The curves are vertically shifted by 0.1 for clarity. The first and second resonance modes are shown (compare to Figure 1 in Ref. \cite{Mikhailov05a}). The position (b) and the amplitude (c) of resonances as a function of $\eta_c$. }
\label{B0a0.8}
\end{figure}

As seen from the above discussion, attaching the contacts to the 2D-EG stripe substantially increases the radiative decay of the 2D plasmon modes even if the retardation parameter of the 2DEG itself is small, and the 2D plasmons in the stripe without contacts could be described in quasi-static approximation. Figure \ref{B0a0.8} shows the absorption spectra of the stripe with contacts at a larger value of $\alpha=0.8$. The overall behavior of modes is similar to that shown in Figure \ref{B0a0.1}, but the modes are substantially broader and the resonance maxima are, respectively, smaller (the quality factor of resonances is lower), as compared to Figure \ref{B0a0.1}. The amplitude of the second mode may exceed that of the first one, due to the same physical reasons that have been discussed in Ref. \cite{Mikhailov05a}. 

At zero magnetic field we have also studied absorption spectra of the system at the parallel polarization of the incident electromagnetic wave ($E_x^0= 0$, $E_y^0\neq 0$). In this case the absorption spectra exhibit a simple behavior with maxima at $\omega=0$ and monotonic decrease of the amplitudes at higher frequencies. No plasmon-related resonances are seen in this case.

\subsection{\label{finiteB}Finite magnetic fields}

\begin{figure*}[ht!]
\includegraphics[width=8.5cm]{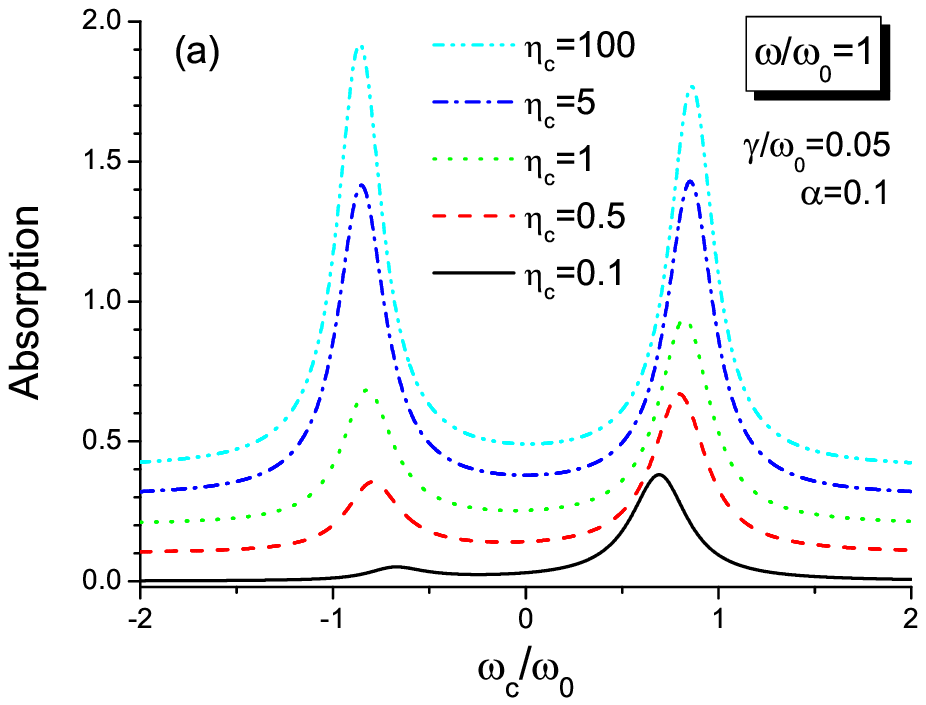}
\includegraphics[width=8.5cm]{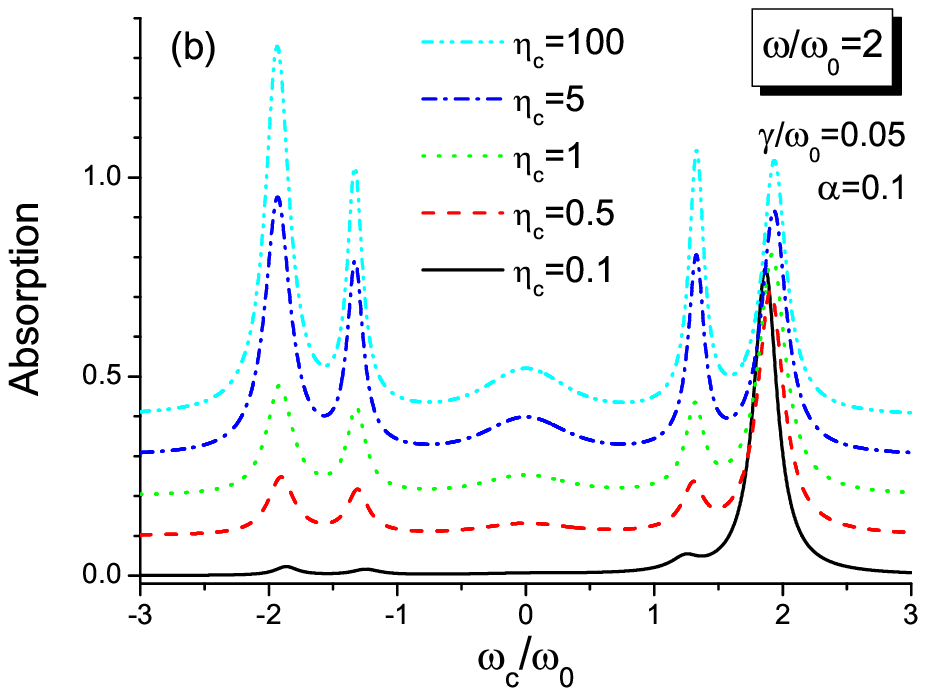}
\caption{(Color online) Magnetic field dependencies of the absorption spectra of a 2D EG stripe placed between contacts at $\gamma/\omega_0=0.05$, $\alpha=0.1$, different contact conductivities, and at frequencies (a) $\omega/\omega_0=1$ (corresponds to $\Omega=0.05$) and (b) $\omega/\omega_0=2$ ($\Omega=0.1$). All curves are vertically shifted by 0.1 for clarity.}
\label{B}
\end{figure*}

Consider the influence of contacts on the absorption spectra of a 2D EG stripe in the presence of magnetic field $B$. Let the incident electromagnetic wave be circularly polarized with $E_y^0/E_x^0=i$ and the magnetic field be both positive and negative, so that in an infinite 2D electron system the radiation would be absorbed only at $B>0$. Figure \ref{B}a shows the absorption spectra of the stripe as a function of magnetic field, at $\omega/\omega_0=1$ and different values of the parameter $\eta_c$. At small $\eta_c$, one sees two resonance maxima at positive and negative $B$ (at $\omega_c/\omega_0\simeq \pm 0.7$). These resonances correspond to excitation of the bulk magnetoplasmons with the frequency $\omega_{mp1}\simeq\sqrt{\omega_{p1}^2+\omega_c^2}$, where $\omega_{p1}$ is the lowest bulk-plasmon frequency at $B=0$. The right resonance, corresponding to the active sense of circular polarization, is much stronger than the left one, as it should be at a small conductivity of contacts (in an infinite system without contacts the left resonance would have a vanishing oscillator strength). When $\eta_c$ increases, the both resonances grow up, as we have seen already in Figure \ref{B0a0.1}b, with the left maximum growing faster than the right one. At $\eta_c\simeq 5$ the two maxima in Figure \ref{B}a become comparable in their height. The two metallic contact wings thus polarize the incident radiation in the gap between them, as a standard wire-grid polarizer (e.g. Ref. \cite{Hecht02}, \S \  8.3). Less expected, however, is that at even higher conductivity of the contacts the left maximum becomes higher than the right one. Such inversion of the absorption maxima can be also seen at other frequencies. Figure \ref{B}b shows the absorption spectra at $\omega/\omega_0=2$. In this case one sees two pairs of bulk magnetoplasmon resonances, with a third resonance appearing at $\omega_c\sim 0$ at the very large conductivity $\eta_c$. The inversion of the right and left absorption maxima can be also seen in this plot.
 
\begin{figure*}[ht!]
\includegraphics[width=8.5cm]{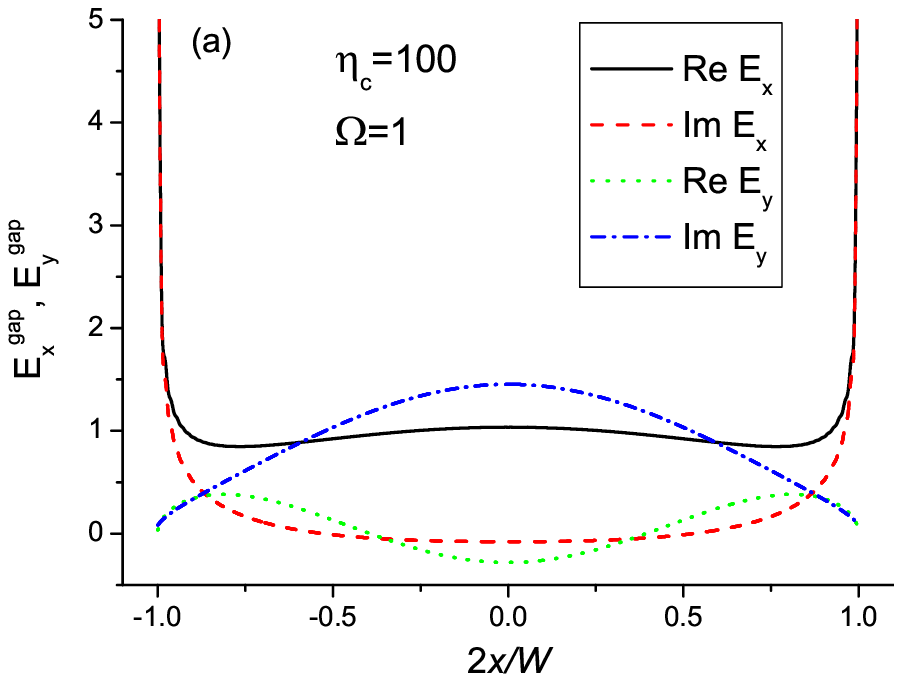}
\includegraphics[width=8.5cm]{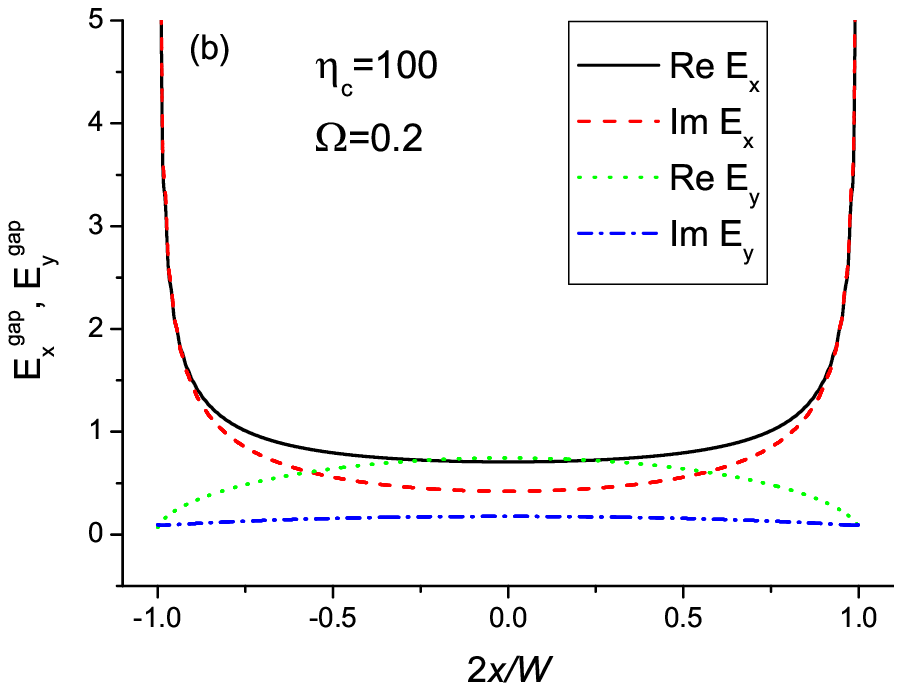}
\caption{(Color online) The electric field inside the gap between the contact wings, in the absence of 2D electrons, at $\eta_c=100$ and (a) $\Omega=1$ and (b) $\Omega=0.2$. The incident-wave electric field was assumed to be circularly polarized with $E_x^0=1$ and $E_y^0=i$. }
\label{Eingap}
\end{figure*}

\begin{figure*}[ht!]
\includegraphics[width=8.5cm]{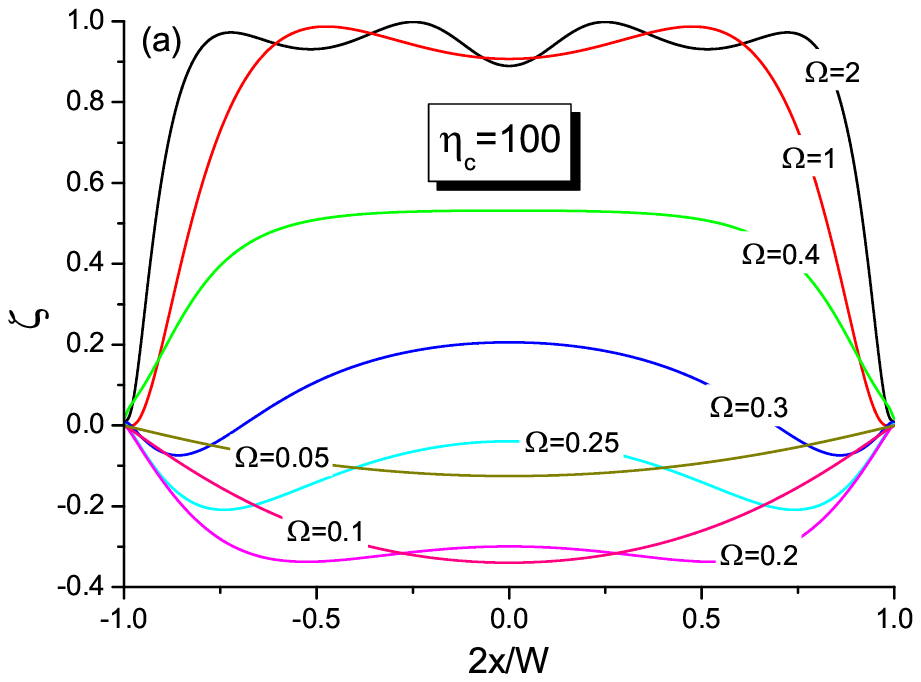}
\includegraphics[width=8.5cm]{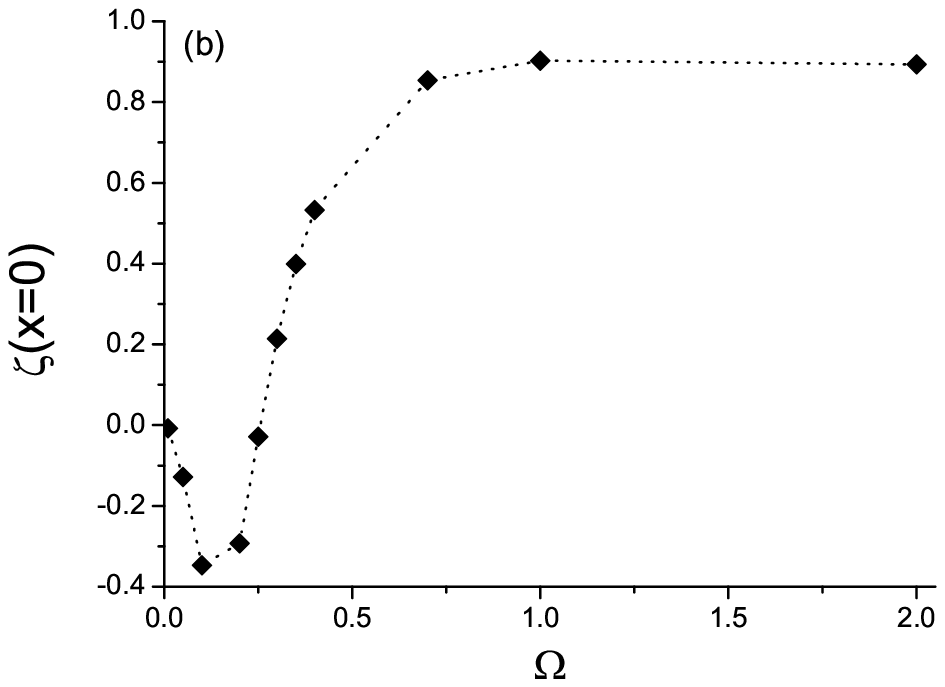}
\caption{(Color online) (a) The degree of the circular polarization $\zeta$ (a) as a function of $x$ at different frequencies $\Omega$, and (b) as a function of $\Omega$ in the center of the gap $x=0$. $\eta_c=100$ on both plots. }
\label{polariz}
\end{figure*}

In order to understand the reason of this effect we calculate the ac electric field ${\bf E}^{gap}
(x,z=0)$ 
in the gap between the contact wings ($|x|<W/2$) {\em in the absence of the 2D electron gas}. In this case equations (\ref{me1})--(\ref{me2}) give
\begin{equation}
\sum_{n=0}{\cal K}_{mn}A_n=\delta_{m0}E_x^0/(1+\eta_c),
\label{me1a}
\end{equation}
\begin{equation}
\sum_{n=0}\left({\cal E}_{mn}-{\cal L}_{mn}\right)B_n=\delta_{m0}E_y^0/(1+\eta_c).
\label{me2a}
\end{equation}
The vectors $A_n$ and $B_n$, calculated from Eqs. (\ref{me1a})--(\ref{me2a}), depend on the contact conductivity $\eta_c$ and the dimensionless frequency $\Omega= \omega W\sqrt{\epsilon}/2\pi c=W/\lambda$. Together with (\ref{Exy}), they describe the field in the gap, screened by the metal contacts but not screened by electrons of the 2D gas; for 2D electrons ${\bf E}^{gap}$ is the external field. The fields $E_x^{gap}(x)$ and $E_y^{gap} (x) $ are complex functions of $x$. Figure \ref{Eingap} shows examples of the dependencies $E^{gap}_{x}(x) $ and $E^{gap}_{y}(x) $ for two different values of $\Omega$. In the gap center, the field $E_x^{gap} $ is close to the field of the incident wave. Near the contacts, $E_x^{gap} $ is much stronger than the field of the incident wave and is linearly polarized, due to the big linear charges induced at the contact edges by the external radiation. The field $E_y^{gap} $ is not so strongly modified in amplitude, but its phase may be quite different at different frequencies $\Omega$. To quantitatively characterize the degree of circular polarization of the local field in the gap, we calculate the quantity
\begin{equation}
\zeta(x)=i\frac{E_xE_y^\star-E_x^\star E_y}{|E_x|^2+|E_y|^2}
\end{equation}
with ${\bf E=E}^{gap}(x,z=0)$. The function $\zeta(x)$ varies between $+1$ and $-1$, with the values $\zeta=\pm 1$ corresponding to the right and left purely circular polarizations and $\zeta=0$ -- to the linearly polarized light. For the incident wave with $E_x^0=1$ and $E_y^0=i$, the $\zeta$ parameter equals $+1$. Figure \ref{polariz}a exhibits $\zeta(x)$ in a gap between good metallic contacts ($\eta_c=100$) and at different frequencies $\Omega$. At $\Omega\gtrsim 1$ (the gap width is larger than the wavelength of radiation $\lambda$) $\zeta(x)$ is close to $+1$ (like the incident wave) everywhere in the gap, except its edges. Near the contacts the local field is linearly polarized. At smaller values of $\Omega$, when $\lambda$ exceeds the gap width $W$, the sense of circular polarization of the local field changes its sign: one sees that at $\Omega\lesssim 0.25$ the function $\zeta(x)$ is negative at all $x$. Physically this is a consequence of screening of the ac electric field in metallic contacts. At very small values of $\Omega$ ($\ll 0.05$) the field becomes linearly polarized ($\zeta\approx 0$) everywhere in the gap. Figure \ref{polariz}b shows $\zeta(x=0)$ at the center of the gap at a function of $\Omega$. One sees that inversion of the circular polarization sense is the case at $\Omega\lesssim 0.25$, or at $W\lesssim \lambda/4$.

\begin{figure*}[ht!]
\includegraphics[width=8.5cm]{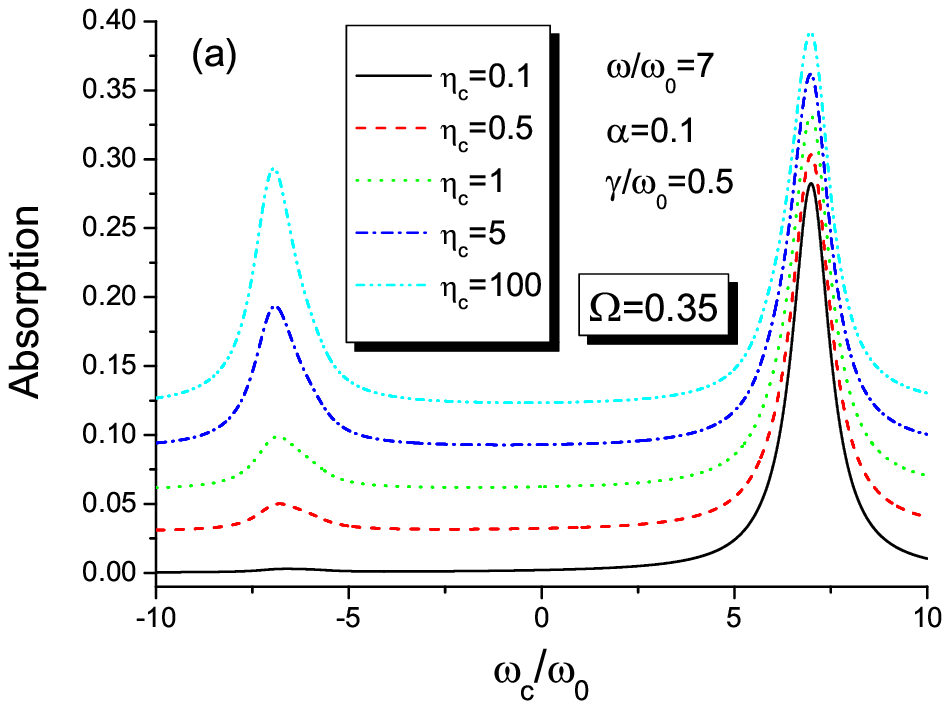}
\includegraphics[width=8.5cm]{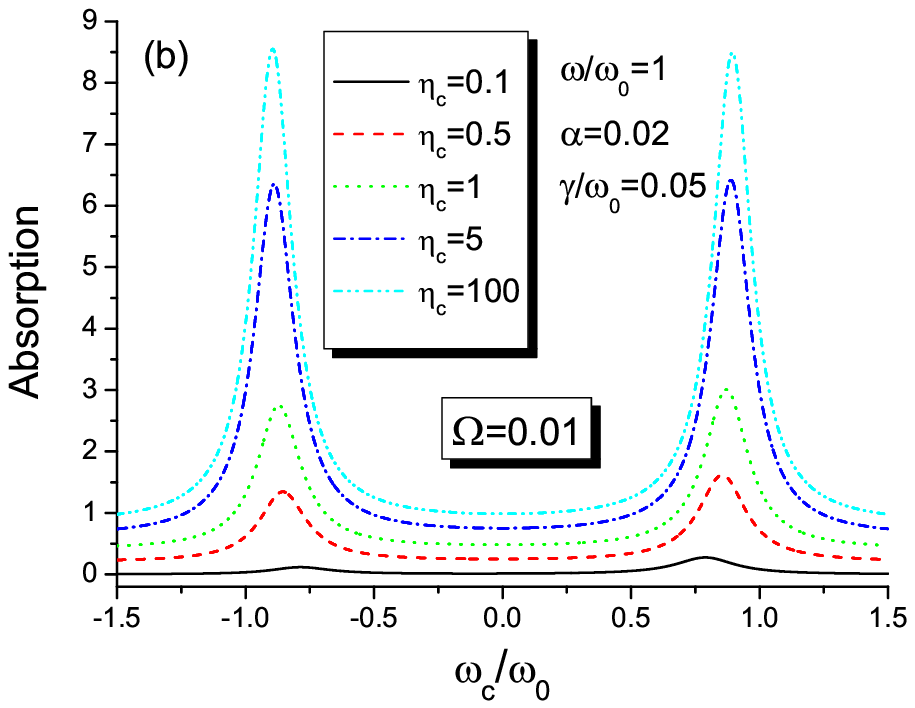}
\caption{(Color online) Absorption spectra of the 2D EG stripe at (a) $\gamma/\omega_0=7$ and $\alpha=0.1$, which corresponds to $\Omega=0.35$ and (b) $\omega/\omega_0=1$ and $\alpha=0.02$, which corresponds to $\Omega=0.01$. All curves are vertically shifted for clarity by 0.03 in (a) and 0.2 in (b). The incident-wave electric field is assumed to be circularly polarized with $E_x^0=1$ and $E_y^0=i$.}
\label{B1}
\end{figure*}

The $B$-dependencies of the absorption spectra shown in Figure \ref{B} correspond to $\Omega=0.05$ and $\Omega=0.1$. As seen from Figure \ref{polariz}, at these frequencies $\zeta$ is negative everywhere in the gap. Therefore the left absorption resonances are higher than the right ones in Figure \ref{B}. As follows from Figure \ref{polariz}, at much smaller, as well as at much bigger values of $\Omega$, the absorption spectra should look ``normally'', with the right maxima, corresponding to the active circular polarization, being higher than or equal to the left ones. Figure \ref{B1}a ($\Omega=0.35$) and Figure \ref{B1}b ($\Omega=0.01$) confirm that, indeed, in these cases the absorption spectra have their usual form. The difference between Figures \ref{B1}a and \ref{B1}b is also clear in view of results shown in Figure \ref{polariz}: at $\Omega=0.35$ (Figure \ref{B1}a) the field in the bulk of the stripe is elliptically polarized with $\zeta\approx +0.4$, therefore the right and the left maxima in Figure \ref{B1}a have a different height. At $\Omega=0.01$ (Figure \ref{B1}b) the field is linearly polarized in the stripe (the plot $\zeta(x)$, not shown in Figure \ref{polariz}a, is very close to zero at all $x$), therefore the right and the left maxima in Figure \ref{B1}b have practically the same height. 

In Ref. \cite{Smet05} the influence of the circular polarization sense on the microwave absorption and the microwave induced zero-resistance states has been experimentally studied. Parameters of the van der Pauw sample in \cite{Smet05} corresponded to the conditions $\omega/\omega_0\gtrsim 10$, $\alpha\simeq 1$, and $\Omega\gtrsim 5$. As seen from Figure \ref{polariz}a, at $\Omega\gg 1$ the circular polarization sense of the field ${\bf E}^{gap}$, acting on 2D electrons, coinsides with that of the incident wave everywhere in the sample, except the very close vicinity of the contacts. In addition, since in the van der Pauw samples the contacts touch the sample edge only in a few small points (in contrast to our model, Figure \ref{geometry}), one should expect a big difference between the right and left cyclotron resonance (CR) absorption peaks. Indeed, the CR line at negative magnetic fields $\omega\approx -\omega_c$ has been found in Ref. \cite{Smet05} to be negligibly small as compared to that at $\omega\approx +\omega_c$ (due to the condition $\omega/\omega_0\gg 1$ the plasmon-related shift of the CR line was negligible in \cite{Smet05}). Thus, the absorption of electromagnetic wave, which is essentially a {\em bulk} effect, was not strongly modified by contacts in \cite{Smet05}. 

In contrast, the giant $1/B$-periodic photoresistance oscillations and the ZRS effect were found in \cite{Smet05} to be notably insensitive to the sense of circular polarization. We believe that this strongly indicates on the {\em edge} or, more exactly, {\em near-contact origin} of these effects. First, the local field near the contacts is linearly polarized, with the electric field, perpendicular to the edge of the system, at any sense of circular polarization of the incident wave, Figures \ref{Eingap} and \ref{polariz}. Second, the field near the edge is {\em much} stronger than that of the incident wave, Figure \ref{Eingap}. Since the photoresistance is a non-linear phenomenon, it will manifest itself, first of all, near the contacts. Third, the near-contact-ZRS-origin hypothesis may resolve the plasma-shift paradox \cite{Mikhailov03c,Mikhailov04a}. Assume that there exist not one (bulk) but two (bulk and near-contact) contributions to the microwave induced photoresistance. The bulk contribution is sensitive to the circular polarization sense, results from the heating of the electron gas by microwave radiation, and leads to weak resonances at the bulk magnetoplasmon frequencies, which have been observed in the earlier microwave photoresistance experiment \cite{Vasiliadou93}, as well as recently in Ref. \cite{Kukushkin06}. In the ZRS experiments \cite{Mani02,Zudov03,Dorozhkin03,Yang03,Mani04,Studenikin04,Willett04,Du04,Smet05} the bulk contribution was not visible because it was masked by a bigger edge (near-contact) oscillating contribution (whose specific nature is not clear at present), which is insensitive to the circular polarization sense and manifests itself only in very-high mobility samples. This hypothesis is consistent with that fact that in some ZRS experiment \cite{Zudov01,Du04} a weak magnetoplasmon-related resonance has been seen on the background of evolving giant photoresistance oscillations.

\section{\label{dc}Summary}

We have theoretically studied the microwave absorption spectra of a finite-width 2D electron-gas stripe, supplied by two semiinfinite side metallic contacts. We have shown that at zero magnetic field the presence of the contacts leads to additional radiative broadening of plasmon-polariton resonances, to the growth of their amplitude (the absorption is growing because the contacts work as antennas), and to the red shift of modes. In finite magnetic field the contacts influence the relative strength of the absorption resonances of the clockwise and counter-clockwise circularly polarized waves. Under certain conditions, the inversion of the resonance absorption strengthes for two types of circular polarized waves is predicted.

It would be interesting to observe the effects predicted in this paper, especially the inversion of the absorption maxima in finite magnetic fields at $W\lesssim \lambda/4$ (Figure \ref{B}), in a direct experiment. The best possible geometry for such an experiment is, probably, a large Corbino disk, as our model is close to it. It would be also interesting to compare absorption spectra in the Corbino disk and in a 2D electron ring \cite{Kovalskii06}.

\begin{appendix}
\section{Estimate of the matrix element ${\cal K}_{00}$ \label{ap1}}

Consider the matrix element 
\begin{equation}
{\cal K}_{00}=2\int_0^\infty \frac{dQ F_0^2(Q)}{1+\frac{i\eta_c}\Omega \sqrt{Q^2-\Omega^2}}
\end{equation}
in the limits $\eta_c\gg 1$ (an ideal metal) and $\Omega\ll 1$. Under these conditions we have
\begin{equation}
{\cal K}_{00}\approx \frac{2\Omega}{i\eta_c}\int_0^\infty \frac{dQ F_0^2(Q)}{\sqrt{Q^2-\Omega^2}}.
\end{equation}
A part of the integral from $Q=0$ up to $Q=|\Omega|$ gives an imaginary contribution and physically corresponds to the radiative decay in the metal (contact) layers. Then we write
\begin{equation}
\frac{i\eta_c}{2\Omega}{\cal K}_{00}=\int_0^{\Omega_0} \frac{dQ F_0^2(Q)}{\sqrt{Q^2-\Omega^2}}+\int_{\Omega_0}^\infty \frac{dQ F_0^2(Q)}{\sqrt{Q^2-\Omega^2}},
\label{K00expand}
\end{equation}
where $\Omega_0$ is an arbitrary number lying in the interval
\begin{equation}
\Omega\ll\Omega_0\ll 1.\label{cond}
\end{equation}

At $Q\ll 1$ the function $F_0^2(Q)=[\sin(\pi Q)/(\pi Q)]^2\approx 1$. Taking into account (\ref{cond}), we can replace $F_0^2(Q)$ by unity in the first integral in (\ref{K00expand}), and ignore $\Omega^2$ as compared to $Q^2$ in the second one. So we get

\begin{eqnarray}
\frac{i\eta_c}{2\Omega}{\cal K}_{00}&=&
\int_0^{\Omega_0} \frac{dQ}{\sqrt{Q^2-\Omega^2}}+
\int_{\Omega_0}^\infty \frac{dQ F_0^2(Q)}{Q} \nonumber \\
&=&
\int_0^{1} \frac{dQ}{\sqrt{Q^2-\Omega^2}}-
\int_{\Omega_0}^1 \frac{dQ}{\sqrt{Q^2-\Omega^2}}+
\int_{\Omega_0}^1 \frac{dQ F_0^2(Q)}{Q}+
\int_{1}^\infty \frac{dQ F_0^2(Q)}{Q}.
\label{K00expand2}
\end{eqnarray}
In the second integral here we can, again, ignore $\Omega^2$ as compared to $Q^2$. Then
\begin{eqnarray}
\frac{i\eta_c}{2\Omega}{\cal K}_{00}&=&
\int_0^{1} \frac{dQ}{\sqrt{Q^2-\Omega^2}}-
\int_{\Omega_0}^1 \frac{dQ}{Q}\left[1-F_0^2(Q)\right]
+
\int_{1}^\infty \frac{dQ F_0^2(Q)}{Q}.
\label{K00expand3}
\end{eqnarray}
Now we can replace the lower limit ($\Omega_0$) in the second integral by zero, as $F_0^2(Q)\approx 1$ at $Q <\Omega_0\ll 1$ and the corresponding error will be negligibly small. Taking the first integral we finally get

\begin{equation}
{\cal K}_{00}\approx
\frac{2\Omega}{i\eta_c}
\ln\frac {2ie^{-C_1}}{\Omega}\approx
\frac{2\Omega}{i\eta_c}
\left(
\ln\frac {0.8 }{\Omega}+i\frac \pi 2
\right),
\label{K00}
\end{equation}
where
\begin{equation}
C_1=\int_{0}^1 \frac{dQ}{Q}\left[1-F_0^2(Q)\right]-
\int_{1}^\infty \frac{dQ }{Q}F_0^2(Q)
=0.915.
\label{C1}
\end{equation}
The formula (\ref{K00}) is valid at $\eta_c\gg 1$ and $\Omega\ll 1$. 

\end{appendix}

\begin{acknowledgments}
We gratefully acknowledge financial support from the Swedish Research Council (Vetenskapsr\aa det), the Swedish Foundation for International Cooperation in Research and Higher Education (STINT), and INTAS. A part of this work has been done during a research visit of one of us (S.M.) to the University of Aizu, Japan; S.M. thanks Professor Victor Ryzhii for warm hospitality and the Japan Society for the Promotion of Science (JSPS) for financial support of this visit. We would also like to thank Igor Kukushkin, Jurgen Smet and Klaus von Klitzing for numerous motivating discussions.  
\end{acknowledgments}

\end{document}